\documentclass[11pt]{article}
\usepackage{comment}
\usepackage[margin=2.5cm]{geometry}
\usepackage{amsmath,amssymb,extarrows,mathtools,graphicx,subfigure,setspace,bbold}
\usepackage{cite}
\usepackage{slashed}
\usepackage[dvipsnames]{xcolor}
\usepackage{hyperref}
\hypersetup{colorlinks=true, linkcolor=blue, citecolor=magenta, linktoc=page}
\numberwithin{equation}{section}
\newcommand{\be}{\begin{equation}}
\newcommand{\bea}{\begin{eqnarray}}
\newcommand{\eea}{\end{eqnarray}}
\newcommand{\ba}{\begin{align}}
\newcommand{\ea}{\end{align}}
\newcommand{\ee}{\end{equation}}

\usepackage{chngcntr}
\counterwithout*{figure}{section}

\begin{document}

\begin{titlepage}
\vspace{10mm}
\begin{flushright}
IPM/P-2018/069 \\

\end{flushright}

\vspace*{20mm}
\begin{center}

{\Large {\bf Subregion Action and  Complexity }\\
}

\vspace*{15mm}
\vspace*{1mm}
{Mohsen Alishahiha${}^\ast$,  Komeil Babaei Velni $^{\dagger}$ and 
M. Reza Mohammadi Mozaffar${}^{\dagger, \ast}$}

 \vspace*{1cm}

{\it ${}^\ast$ School of Physics,
Institute for Research in Fundamental Sciences (IPM)\\
P.O. Box 19395-5531, Tehran, Iran\\
 ${}^\dagger$ Department of Physics, University of Guilan,
P.O. Box 41335-1914, Rasht, Iran\\
}

 \vspace*{0.5cm}
{E-mails: {\tt alishah@ipm.ir, babaeivelni@guilan.ac.ir, mmohammadi@guilan.ac.ir}}%

\vspace*{1cm}
\end{center}

\begin{abstract}
We evaluate  finite part of the on-shell action  for black brane  solutions of Einstein 
gravity on different subregions 
of spacetime  enclosed by null boundaries.  These subregions include the intersection
of WDW patch with past/future interior and left/right exterior for a two sided black brane. 
Identifying the on-shell action on the exterior regions with subregion complexity one finds that it obeys subadditivity condition. This gives an insight to define a new quantity named mutual complexity.
We will also consider certain subregion that is a part of spacetime which could be causally
connected to an operator localized behind/outside  the horizon. 
Taking into account all terms
needed to have a diffeomorphism invariant action with a well-defined variational principle, one
observes  that the main contribution that results to a nontrivial behavior of the on-shell action
comes from joint points where two lightlike boundaries (including horizon) intersect.   
A spacelike boundary gives rise to a linear time growth, while we have 
a classical contribution due to a timelike boundary that is given by the free energy.

\end{abstract}

\end{titlepage}

\newpage
\tableofcontents
\noindent
\hrulefill
\onehalfspacing

\section{Introduction}

Based on earlier works of  \cite{{Susskind:2014rva},{Stanford:2014jda} } it was conjectured that 
computational complexity associated  with a boundary state may  be identified with 
the on-shell action evaluated on a certain subregion of the bulk spacetime
 \cite{{Brown:2015bva},{Brown:2015lvg}}. The corresponding 
subregion is Wheeler-DeWitt (WDW) patch of the spacetime that is  the domain of 
dependence of any Cauchy 
surface in the bulk whose intersection with the asymptotic boundary is the time slice 
on which the state is defined.

This proposal, known as ``complexity equals action'' (CA), has been used to explore several
properties of computational  complexity for those field theories that have gravitational dual
\footnote{We would like to stress that on the gravity side there is another proposal for computing 
the computational complexity, known as ``complexity equals volume'' (CV)
\cite{{Susskind:2014rva},{Stanford:2014jda} }. The generalization of CV proposal 
to subsystems has been done in 
\cite{Alishahiha:2015rta} (see also \cite{{Ben-Ami:2016qex},{Couch:2016exn},{Carmi:2016wjl},
{Abt:2017pmf},{Abt:2018ywl},{Bakhshaei:2017qud},{Chen:2018mcc},{Ageev:2018nye}}). 
Yet another approach to complexity based on Euclidean path-integral 
has been introduced in \cite{{Miyaji:2016mxg},{Caputa:2017urj},{Caputa:2017yrh}}.
For a recent development and its possible  relation with CA approach see 
\cite{Takayanagi:2018pml}.}.
In particular the growth rate of complexity has been studied for an eternal  black hole in
\cite{Carmi:2017jqz}.
It was shown that although in the late time the growth rate  approaches a constant value that is 
twice of the mass of the black hole, the constant is approached from above, violating the Lloyd's 
bound \cite{Lloyd:2000}. Of course this is not the case for a state followed by a global quench
\cite{Moosa:2017yvt}. It is worth to mention that recently there has been some progress for
studying the computational complexity of a state in field theory\cite{Hashimoto:2017fga,Jefferson:2017sdb,Chapman:2017rqy,Yang:2017nfn,Khan:2018rzm,Yang:2018nda,Hackl:2018ptj,Alves:2018qfv,Bhattacharyya:2018wym,Guo:2018kzl,Bhattacharyya:2018bbv,Yang:2018tpo}. 

So far the main concern in the literature was the growth rate of complexity and therefore
the on-shell action was computed up to time independent terms \cite{Kim:2017qrq,Moosa:2017yiz,Swingle:2017zcd}. Moreover it was also
shown that the time dependent effects are controlled by the regions behind the horizon. 
We note, however, that in order to understand holographic complexity better it is crucial to have 
the full expression of it. It is also important to evaluate the  contribution of different 
parts (inside and outside of the horizon)  of the WDW patch, specialty. It is also illustrative to compute on-shell action on 
a given subregion of spacetime enclosed by null boundaries, that is not  necessarily the WDW 
patch. Indeed this is one of the aim of the  present work to carry out these computations 
explicitly. Moreover we will carefully identify the contribution of each term in the action. 

Since we are interested in the on-shell action, it is crucial to make clear what one means by 
``on-shell action''. In general an action could have several terms that might be important due to certain physical reason. In particular in order to have a well-defined variational principle 
with Dirichlet boundary condition one needs to add certain Gibbons-Hawking-York boundary 
terms at spacelike and timelike boundaries\cite{York:1972sj,Gibbons:1976ue}. Moreover to accommodate null boundaries it 
is also crucial to add the corresponding boundary terms on the null boundaries  as well as 
certain joint  action at points  where a null boundary  intersects to other  boundaries 
\cite{{Parattu:2015gga},{Lehner:2016vdi}}.
  
Restricted to Einstein gravity and assuming  to have a well-defined variational principle 
one arrives at the following action\cite{Lehner:2016vdi}
\bea\label{ACT0}
I^{(0)}&=&\frac{1}{16\pi G_N}\int d^{d+2}x\sqrt{-g}(R-2\Lambda)+\frac{1}{8\pi G_N}
\int_{\Sigma^{d+1}_t} K_t\; d\Sigma_t\cr &&\cr
&& \pm\frac{1}{8\pi G_N} \int_{\Sigma^{d+1}_s} K_s\; d\Sigma_s\pm \frac{1}{8\pi G_N} 
\int_{\Sigma^{d+1}_n} K_n\; dS 
d\lambda \pm\frac{1}{8\pi G_N} \int_{J^d} a\; dS\,.
\eea
Here the timelike, spacelike, and null boundaries and also joint points are denoted by $
\Sigma_t^{d+1}, \Sigma_s^{d+1}, \Sigma_n^{d+1}$ and $J^d$, respectively. The extrinsic 
curvature of the corresponding boundaries are given by $K_t, K_s$ and $K_n$. The function $a$ 
at the intersection of the boundaries is given by the logarithm of the inner product of the 
corresponding normal vectors and $\lambda$ is the null coordinate defined on 
the null segments. The sign of different terms depends on the relative position 
of the boundaries and the bulk region of interest (see \cite{Lehner:2016vdi} for more details).

As far as the variational principle is concerned the above action defines a consistent 
theory. Nonetheless one still has possibilities to add certain boundary terms that do not 
alter the boundary condition, but could have nontrivial contribution to the on-shell action. 
Therefore it is important to fix these terms using certain physical principles before computing 
the on-shell action.

In particular one can see that the above action is not invariant under a reparametrization of the
null generators. Therefore one may conclude that the above action does not really
define a consistent theory. Actually to maintain the invariance under a reparametrization of the null
generators one needs to add an extra term to the action as follows\cite{Lehner:2016vdi}
(see also \cite{Reynolds:2016rvl})\footnote{ The importance of this term has also been
 emphasized in \cite{{Alishahiha:2018tep},{Chapman:2018dem},{Chapman:2018lsv}}
 where it was shown that it is essential to consider the contribution of this term to the 
 complexity.}
\be\label{AMB}
I^{\rm amb}=\frac{1}{8\pi G_N}\int_{\Sigma_n^{d+1}} d^dxd\lambda\,\sqrt{\gamma}\,\Theta\,
\log\frac{|\tilde{L}\Theta|}{d},
\ee
where $\tilde{L}$ is an undetermined  length scale and  
$\gamma$ is the determinant of the induced 
metric on the joint point where two null segments intersect, and 
\be
\Theta=\frac{1}{\sqrt{\gamma}}\frac{\partial\sqrt{\gamma}}{\partial\lambda}.
\ee 
Although even with this extra term the length scale $\tilde{L}$ remains undetermined, adding 
this term to action \eqref{ACT0} would define a consistent theory. Therefore in what follows
by evaluating ``on-shell action'' we mean to consider $I=I^{(0)}+I^{\rm amb}$.
We note, however, that the resultant on-shell action may or may not be UV finite. Thus,
one may want  to get finite on-shell action (as we do for gravitational free energy)
that requires to add certain counterterms. Actually these terms are also required from
holographic renormalization (see {\it e.g.} \cite{Skenderis:2002wp}).  Of course in this 
paper we will not consider such counterterms and those needed due to null boundaries
\cite{TBA}.

The aim of this article is to compute on-shell action on certain subregions 
behind and outside the horizon enclosed by null boundaries.  We will consider an 
eternal black brane that provides a gravitational dual for a thermofield double 
state. Those subregions that are behind the horizon are UV finite and time dependent, 
though those outside the horizon are typically  UV divergent and time independent. 

To proceed we will consider a $(d+2)$-dimensional black brane solution in Einstein gravity whose 
metric is\footnote{Due to flat boundary of the  black brane solution, we will be able to present 
our results in simple compact forms. } 
\be\label{metric}
ds^2=\frac{L^2}{r^{2}}
\left(-{f(r)}dt^2
+\frac{dr^2}{f(r)}+\sum_{i=1}^d d\vec{x}^2\right),\;\;\;\;\;\;\;\;f(r)=1-\left(\frac{r}{r_h}\right)^{d+1},
\ee
where $r_h$ is the radius of horizon and $L$ denotes the AdS radius. In terms of these 
parameters the entropy, mass and Hawking temperature of the corresponding black brane are 
\be
S_{th}=\frac{V_d L^d}{4G_Nr_h^{d}},\;\;\;\;\;\;\;\;\;\;M=\frac{V_dL^d}{16\pi G_N}\frac{d}
{r_h^{d+1}},\;\;\;\;\;\;\;\;\;\;\;T=\frac{d+1}{4\pi r_h},
\ee
with  $V_d$ being the volume of $d$-dimensional internal space of the metric parametrized by 
$x_i, \,i=1,\cdots d$. It is also useful to note that
 \be\label{bulk}
\sqrt{-g}\left(R-2\Lambda\right)=-2(d+1)
\frac{L^d}{r^{d+2}}.
\ee
The organization of the paper is as follows. In the next section we will consider 
on-shell action on the WDW patch that using  CA proposal may be related to 
the holographic complexity of the dual state. Our main concern is to present a closed form 
for the on-shell action.  We will also compute on-shell action for past patch that is obtained 
by continuing the past null boundaries all the way to the past singularity. We will also
compute on-shell action on the intersection of  WDW patch with past and future interiors. 
We study the time evolution of holographic uncomplexity too. 
In section three we will consider different patches that are outside the horizon. This includes 
the intersection of WDW patch with entanglement wedge that could thought of as
CA subregion complexity. The last 
section is devoted to discussion and conclusion where we present the interpretation 
of our results.    


\section{Complexity and Subregions Behind the Horizon}

\subsection{CA Proposal}

\begin{figure}
\begin{center}
\includegraphics[scale=0.85]{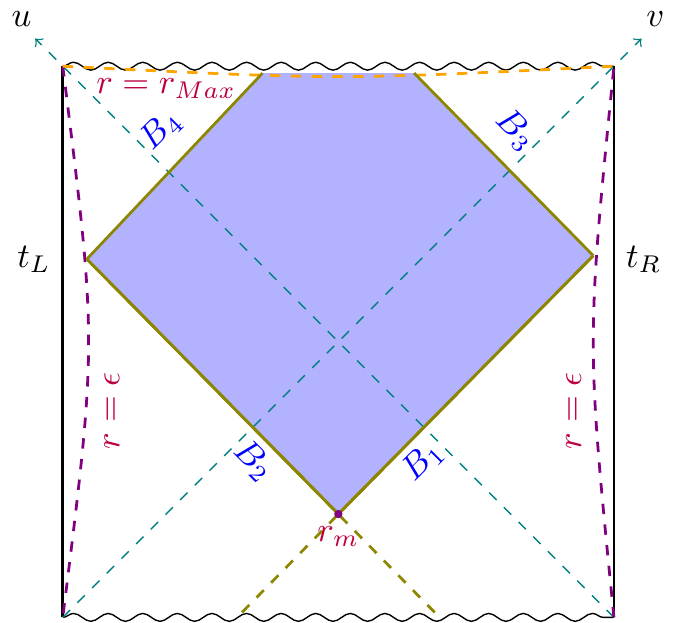}
\hspace{1.5cm}
\includegraphics[scale=0.85]{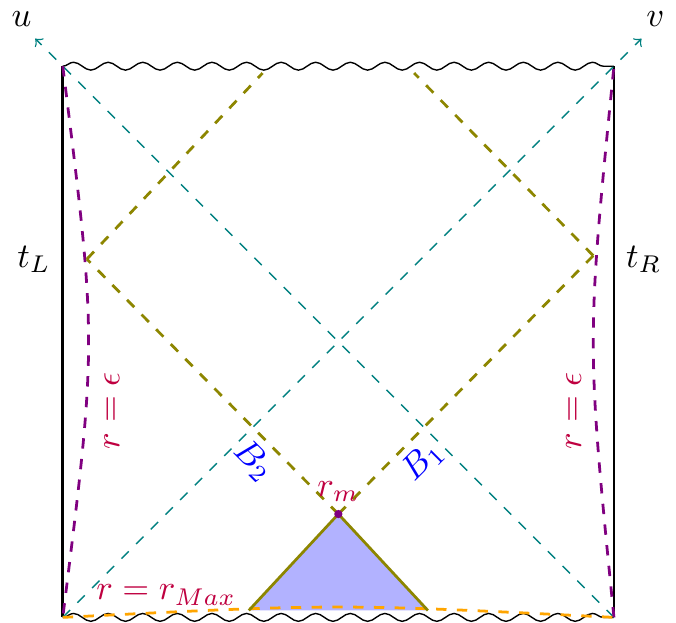}
\end{center}
\caption{Penrose diagram of the WDW patch of an eternal AdS black hole assuming $t_R=t_L$. 
\textit{Left}: WDW patch on which the on-shell action is computed to find the complexity. 
\textit{Right}: Past patch corresponding to the WDW patch.
 The past patch may be identified as a part that is casually connected 
 to an operator localized at $r=r_m$ behind the horizon.}
\label{WDWP}
\end{figure}

In this section using CA proposal we would like to evaluate complexity for the eternal two sided
black brane which is dual to the thermofield double state in the boundary theory. 
Holographically one should compute on-shell action on WDW patch as depicted in the 
left panel of  figure \ref{WDWP}. Using the symmetry of the  Penrose diagram of the 
eternal black hole, we shall consider a symmetric 
configuration with times $t_R=t_L=\frac{\tau}{2}$. Actually  this question  has been 
already addressed \cite{Carmi:2017jqz} where the full  time dependence of complexity 
has been obtained where it was  shown that the holographic complexity  violates the Lloyd's bound in this case\footnote{The  one sided black hole was also discussed in \cite{Moosa:2017yvt,Alishahiha:2018tep,Chapman:2018dem,Chapman:2018lsv} where it confirmed that 
in this case the Lloyd's bound is respected.}.  Of course our main interest in the present 
paper is to study the finite part of the on-shell action. In this subsection we will present the results and computations rather in details. Due to the similarity of computations, in the rest of the paper the computations will be a little bit brief.

To proceed we note that the null boundaries of the corresponding WDW patch are 
(see left panel of figure \ref{WDWP}) 
\bea
&&B_1:\,\,t=t_R-r^*(\epsilon)+r^*(r),\;\;\;\;\;\;\;\;\;\;B_2:\,\,t=-t_L+r^*(\epsilon)-r^*(r),\cr &&\cr
&&B_3:\,\,t=t_R+r^*(\epsilon)-r^*(r),\;\;\;\;\;\;\;\;\;\;B_4:\,\,t=-t_L-r^*(\epsilon)+r^*(r),
\eea
by which the position of the joint point $m$  is given by\footnote{Note that in our notation one has
 $r^*(r)\leq 0$.}
\be\label{rm}
\tau\equiv t_L+t_R=2(r^*(\epsilon)-r^*(r_m)).
\ee

Let us now compute the on-shell action over the corresponding WDW patch. As we already
mentioned the action consists of several parts that include bulk, boundaries  and joint
actions. Using equation \eqref{bulk} the bulk action is\cite{Carmi:2017jqz}
\be
I^{\rm bulk}_{\rm WDW}=-\frac{V_dL^d}{4\pi G_N}(d+1)\left(2\int_\epsilon^{r_{Max}} 
\frac{dr}{r^{d+2}}
(r^*(\epsilon)-r^*(r)) +\int_{r_m}^{r_{Max}} \frac{dr}{r^{d+2}} 
 (\frac{\tau}{2}-r^*(\epsilon)+r^*(r))\right)\,.
\ee
By making use of an integration by parts the above bulk action reads
\bea
I^{\rm bulk}_{\rm WDW}&=&-\frac{V_dL^d}{4\pi G_N}\left(2\int_\epsilon^{r_{Max}} 
\frac{dr}{r^{d+1}f(r)}
-\int_{r_m}^{r_{Max}} \frac{dr}{r^{d+1}f(r)} \right)\cr &&\cr
&=&-\frac{V_dL^d}{4\pi G_N}\left(\frac{\tau+\tau_c}{2r_h^{d+1}}
+\frac{2}{d\, \epsilon^d}-\frac{1}{d\,r_m^d}\right)\,,
\eea
where $\tau_c=2(r^*(\epsilon)-r^*(r_{Max}))$ is the critical time below which the time derivative 
of complexity vanishes \cite{{Brown:2015lvg},{Carmi:2017jqz}}. More explicitly one has (see
also \cite{Carmi:2017jqz})
\be
\tau_c=\frac{1}{2T}\,\frac{1}{\sin\frac{\pi}{d+1}}.
\ee 

To find the boundary contributions we note that using the affine parametrization for 
the null directions, the corresponding boundary terms vanish\footnote{For affine 
parametrization of the null direction, the extrinsic curvature of the null boundary will be 
zero and therefore there is no contribution from null boundaries. In this paper we  always use 
this parametrization and therefore we do not need to consider the boundary terms for 
null boundaries.}  
 and we are left with just 
a spacelike boundary at future singularity whose contribution is given by
\be\label{surf1}
I^{\rm surf}_{\rm WDW}=-\frac{1}{8\pi G_N}\int d^dx\, \int_{-t_L-r^*(\epsilon)+r^*(r)}^{
t_R+r^*(\epsilon)-r^*(r)} dt\, \sqrt{h}K_s 
\Big|_{r=r_{\rm Max}},
\ee
where $K_s$ is the the trace of extrinsic curvature of the boundary at $r=r_{\rm Max}$ and $h$ 
is the determinant of the induced metric. To compute this term it is 
useful to note that for a constant $r$ surface using the metric \eqref{metric} one has
\be
\sqrt{h} K=-\sqrt{g^{rr}}\partial_r\sqrt{h}=-\frac{1}{2}\frac{L^d}{r^{d}}\left(\partial_r f(r)-
\frac{2(d+1)}{r}f(r)\right).
\ee
Plugging the above expression into \eqref{surf1} and evaluating it at $r=r_{\rm Max}$ one 
finds  
\bea\label{SURF}
I^{\rm surf}_{\rm WDW}=\frac{V_dL^d}{8\pi G_N }(d+1)\,\frac{\tau+\tau_c}{2r_h^{d+1}}\,.
\eea

There are also several joint points which may contribute to the on-shell action. Two of them 
are  located at the future singularity that  have zero contributions, while the contributions
of the three remaining points at $r=\epsilon$ and $r=r_m$ are given by
\bea
I^{\rm joint}_{\rm WDW}=2\times \frac{-1}{8\pi G_N}\int_{\epsilon} d^dx \sqrt{\gamma} \;
\log \frac{\left|{k_1\cdot k_2}\right|}{2}
+\frac{1}{8\pi G_N}\int_{r_m} d^dx \sqrt{\gamma} \;\log \frac{ \left|k_1\cdot k_2\right|}{2}\,,
\eea
where the factor of 2 is due to the two joint points at $r=\epsilon$ for  left and right boundaries. 
Here $k_1$ and $k_2$ are the null vectors associated with the null boundaries
\be
k_1={\alpha}\left(-dt+
\frac{dr}{f(r)}\right),\;\;
\;\;\;\;k_2={\beta}\left(dt+
\frac{dr}{f(r)}
\right)\,.
\ee
Here  $\alpha$ and $\beta$ are two constants appearing due to the ambiguity of 
the normalization of normal vectors of null segments. Therefore one gets
\be
I^{\rm joint}_{\rm WDW}=
-\frac{V_dL^d}{4\pi G_N}\,\frac{\log\frac{\alpha\beta\epsilon^2}{L^2}}{\epsilon^d}
+\frac{V_dL^d}{8\pi G_N}\left(\frac{
\log \frac{\alpha\beta r_m^2}{L^2}}{r_m^d}-\frac{\log |f(r_m)|}{r_m^d}\right).
\ee
It is clear from the above expression that the result suffers from an ambiguity associated 
with the normalization of null vectors. This ambiguity may be fixed either by fixing the constants
$\alpha$ and $\beta$ by hand or adding a proper term to the action.  Actually as we have already 
mentioned in order to maintain the diffeomorphism invariance of the action 
we will have to add another term given by equation \eqref{AMB}.
 Note that even with this term we are still left with 
an undetermined free parameter.  In the present case taking into account 
all four null boundaries one gets\footnote{ Note that for boundary associated with $k_1$ one has 
$\frac{dr}{d\lambda}=\alpha\frac{r^2}{L^2}$ and $\Theta=2d\alpha\frac{r}{L^2}$. For the other 
null vector one should replace $\alpha$ with $\beta$.} 
\be
I^{\rm amb}_{\rm WDW}=-\frac{V_dL^d}{8\pi G_N}\left(\frac{\log\frac{\alpha\beta 
\tilde{L}^2 r_m^2}{{L}^4}}
{r_m^{d}}+
\frac{2}
{d\,r_m^{d}}\right)+\frac{V_dL^d}{4\pi G_N}\left(\frac{\log\frac{\alpha\beta \tilde{L}^2\epsilon^2}{L^4}}
{\epsilon^{d}}+
\frac{2}
{d\,\epsilon^{d}}\right).
\ee
Now we have all terms in the action evaluated on the WDW patch. Therefore one arrives
at 
\bea\label{com}
I_{\rm WDW}&=&I^{\rm bulk}+I^{\rm surf}+I^{\rm joint}+I^{\rm amb}\cr &&\cr
&=&\frac{V_dL^d}{8\pi G_N}\left[
\frac{2}
{\epsilon^{d}}
\log\frac{\tilde{L}^2}{L^2}+\frac{d-1}
{2r_h^{d+1}}(\tau+\tau_c)-
\frac{\log \frac{\tilde{ L}^2|f(r_m)|}{L^2}}{r_m^d}\right].
\eea
It is important to note that in order to have a meaningful result the divergent term should be 
positive that is the case for $\tilde{L}\geq L$. On the other hand  setting $\tilde{L}=L$ the
divergent term will drop and one gets a finite result consisting of 
two contributions\footnote{Actually in the context of holographic renormalization one 
would add certain 
counterterms to make on-shell action finite. In the present case  to remove the divergent term 
one may add a counterterm in the following form
$$I^{\rm ct}=\frac{1}{8\pi G_N}\int d\lambda d^dx\,\sqrt{\gamma}\,\Theta \log
\frac{L^2}{\tilde{L}^2},$$ which is essentially equivalent  to set $\tilde{L}=L$ and then we are 
left with finite on-shell action. Of course in this paper we keep the 
length scale $\tilde{L}$ undetermined.}: one from  
the future spacelike singularity and a contribution from the joint point at $r=r_m$ given as follows
\bea
I_{\rm WDW}=\frac{V_dL^d}{8\pi G_N}\left[\frac{d-1}
{2r_h^{d+1}}(\tau+\tau_c)-
\frac{\log |f(r_m)|}{r_m^d}\right].
\eea
 It is also
interesting to note that for $r_m\rightarrow r_{Max}$ where $\tau\rightarrow \tau_c$ one gets
\bea
I_{\rm WDW}=\frac{V_dL^d}{8\pi G_N}\frac{d-1}
{r_h^{d+1}}\tau_c=\frac{d-1}{d+1}\,\frac{S_{th}}{\sin\frac{\pi}{d-1}},
\eea
which is identically zero for $d=1$. This might be thought of as complexity of formation of the 
black brane \cite{Chapman:2016hwi}. On the other hand, using the fact 
that $\log |f(r_m)|\approx -\frac{(d+1)\tau}{2r_h}$
for $r_m\rightarrow r_h$ (see next section), one gets linear growth at late times
\bea
I_{\rm WDW}\approx \frac{V_dL^d}{8\pi G_N}\frac{d}
{r_h^{d+1}}\tau=2M\tau\,,
\eea
as expected.


\subsection{Past Patch}

In this subsection we would like to compute on-shell action on the past patch defined by 
the colored triangle shown in the right panel of figure \ref{WDWP}.  Clearly the rate of change 
of on-shell action on the past patch is the same as that of the WDW patch. 
Another way to think of the past patch is to consider an operator localized at $r=r_m$ 
behind the horizon. The part of spacetime  that can be causally connected to the operator is the
triangle depicted in figure \ref{WDWP}. Following CA proposal one may think of the on-shell action 
evaluated on the past patch as the complexity associated with the operator. 

Let us compute the on-shell action for the past patch. To proceed we note that using 
the notation of the 
previous subsection the contribution of the bulk term to the on-shell action is 
\bea
 I^{\rm bulk}_{\rm past}&=&-\frac{V_dL^d}{8\pi G_N}(d+1)\int^{r_{Max}}_{r_m} 
\frac{dr}{r^{d+2}}\int^{t_R-r^*(0)+r^*(r)}_{-t_L+r^*(0)-r^*(r)}dt\cr &&\cr
&=&\frac{V_dL^d}{4\pi
G_N}\int_{r_m}^{r_{Max}}\frac{dr}{r^{d+1}f}=\frac{V_dL^d}{4\pi
G_N}\left(\frac{\tau-\tau_c}{2r_h^{d+1}}+\frac{1}{d\,r_m^{d}}\right).
\eea
Here to get the second line we have performed an integration by parts. On the other hand the 
contribution of the spacelike  boundary at past singularity is found to be
\bea
I^{\rm surf}_{\rm past}=
\frac{1}{8\pi G_N}\int d^dx\int^{t_R-r^*(0)+r^*(r)}_{-t_L+r^*(0)-r^*(r)}dt \sqrt{h}K_s\bigg|_{r=
r_{\rm Max}}
=\frac{V_dL^d}{8\pi G_N }
(d+1)\frac{\tau-\tau_c}{2r_h^{d+1}}\,.
\eea
There are also three  joint points two of which at $r=r_{\rm Max}$ and one at $r=r_m$. The 
corresponding contributions to the on-shell 
action for those at $r_{\rm Max}$ vanish for large $r_{\rm Max}$, while the contribution of 
that at $r=r_m$ is given by
\be 
I^{\rm joint}_{\rm past}
=\frac{1}{8\pi G_N}\int d^{d-1}x \sqrt{\gamma} \;\log \frac{\left|{k_1\cdot k_2}\right|}{2}
=\frac{V_dL^d}{8\pi G_N}\left(\frac{
\log {\frac{\alpha\beta r_m^2}{L^2}}}{r_m^{d}}-\frac{\log |f(r_m)|}{r_m^{d}}\right).
\ee
Finally  the contribution of term needed to remove the ambiguity is 
\be
I^{\rm amb}_{\rm past}
=-\frac{V_dL^d}{8\pi G_N}\left(\frac{\log\frac{\alpha\beta \tilde{L}^2r_m^2}{L^4}}
{r_m^{d}}+
\frac{2}
{d\,r_m^{d}}\right).
\ee
Therefore altogether one arrives at
\be
I_{\rm past}=\frac{V_dL^d}{8\pi
G_N}\left(\frac{d-1}{2r_h^{d+1}}(\tau-\tau_c)
-\frac{\log\frac{\tilde{L}^2 |f(r_m)|}{L^2}}{r_m^{d}}\right),
\ee
which is UV finite even with arbitrary finite length scale $\tilde{L}$. Note that for $r_m\rightarrow
r_{\rm Max}$ where $\tau\rightarrow \tau_c$ the on-shell action for past patch vanishes 
identically. On the other hand, in the late times where $r_m\rightarrow r_h$ one finds linear 
growth as expected.
 

\subsection{Intersection of WDW Patch with Past and Future Interiors}

Even for a static geometry, such as eternal black hole, the interior of black hole grows
with time indicating that there could be a quantity in the dual field theory that grows with 
time far after the system reaches the thermal equilibrium. Indeed this was the 
original motivation for holographic computational complexity to be identified with the volume of the 
black hole interior. 

\begin{figure}
\begin{center}
\includegraphics[scale=0.9]{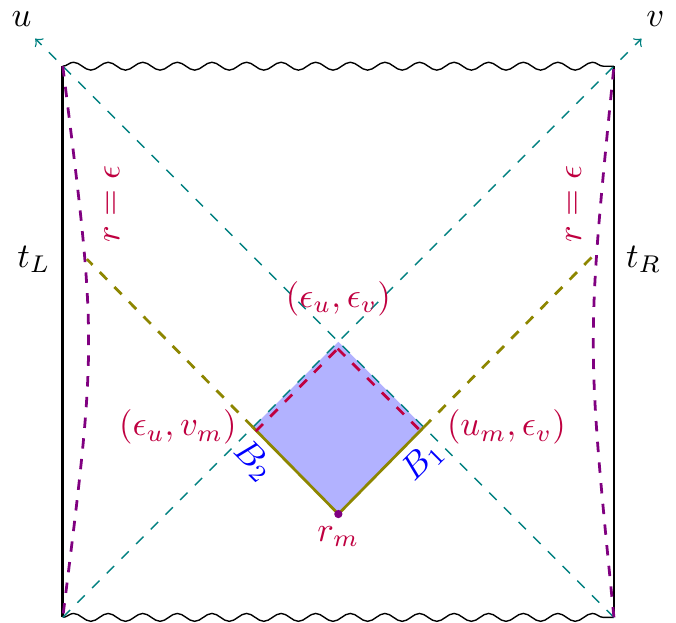}
\includegraphics[scale=0.9]{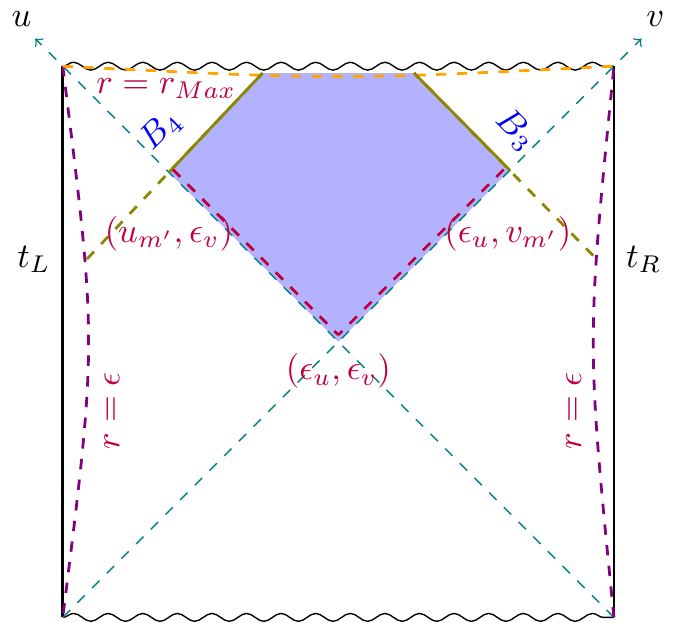}
\end{center}
\caption{{\it Left:} Intersection of WDW patch with the past interior. {\it Right:} Intersection 
of WDW patch with the future interior. }
\label{fig:E}
\end{figure}

In the previous subsection we have computed the on-shell action 
over whole WDW patch. The aim of this subsection is to compute on-shell action 
in the intersection of WDW patch with black brane interior. This consists of past and future
interiors as shown in figure \ref{fig:E}. Actually these subregions are the main 
parts that contribute to the time dependence of complexity of the dual state. It is, however, 
instructive to study these parts separately.\footnote{On-shell action for subregion 
in the black hole interior has been also studied in  \cite{Barbon:2018mxk} where
it was argued that complexity may be used as a probe to study the nature of 
different singularities.} 

\subsubsection{Past Interior} 

To begin with we first  consider the intersection of WDW patch with past interior as shown 
in the left panel of figure \ref{fig:E}.  Actually  one may use the results of the previous
 subsection to write different terms contributing to the on-shell action. To start with we 
 note that for the bulk term one has
\bea
I^{\rm bulk}_{\rm PI}=\frac{V_{d}L^d}{4\pi G_N}(d+1)\int_{r_h}^{r_m} \frac{dr}{r^{d+2}}
\left(\frac{\tau}{2}-r^*(\epsilon)+r^*(r)\right)=\frac{V_{d}L^d}{4\pi G_N}\left(\frac{1}{d\,r_m^d}-
\frac{1}{d\, r_h^d}\right).
\eea

There are four joint points, one at $r=r_m$ and three at $r=r_h$ that contribute to the 
on-shell action. It is, however, important  to note that  those points at the  horizon are not at 
the same point. In other words the radial coordinate $r$ is not suitable to make a 
distinction between these points.  Indeed 
to distinguish between these points, following \cite{Agon:2018zso}, it is convenient to use the
following  coordinate system  for the past interior 
\bea
u=-e^{-\frac{1}{2}f'(r_h)(r^*(r)-t)},\;\;\;\;\;\;\;\;\;\;\;\; v=-e^{-\frac{1}{2}f'(r_h)(r^*(r)+t)}\,.
\eea
In this coordinate system the horizon is located 
at $uv=0$ ({\it i.e.} $r^*(r_h)= -\infty$). This equation has three nontrivial
solutions given by $(u=0,v\neq 0)$, $(u\neq 0,v=0)$ and $(u=0,v= 0)$ that correspond to three
joint points at the horizon shown in figure \ref{fig:E}. Since both $r^*(r)$  and 
$\log f(r)$ are singular at $r=r_h$, one may regularize the contribution of these three points  
by setting the horizon at $v=\epsilon_v$ and $u=\epsilon_u$. In this notation the joint points are 
given by $(\epsilon_u,v_m)$, $(u_m,\epsilon_v)$ and $(\epsilon_u,\epsilon_v)$
as depicted in figure \ref{fig:E}. In what follows the radial coordinate associated with these three
points are denoted by $r_{v_m}, r_{u_m}$ and $r_\epsilon$, respectively. Using this notation the contribution of joint points is 
\bea\label{joint-2} 
I^{\rm joint}_{\rm PI}&=&\frac{V_d L^d}{8\pi G_{N}}
\left(\frac{\log\frac{\alpha\beta r_m^2}{L^2|f(r_m)|}}{r_m^{d}}
-\frac{\log\frac{\alpha\beta r_{u_m}^2}{L^2|f(r_{u_m})|}}{r_{u_m}^{d}}
-\frac{\log\frac{\alpha\beta r_{v_m}^2}{L^2|f(r_{v_m})|}}{r_{v_m}^{d}}
+\frac{\log\frac{\alpha\beta r_\epsilon^2}{L^2|f(r_\epsilon)|}}{r_\epsilon^{d}}\right)\\&&\cr
&=&-\frac{V_d L^d}{8\pi G_{N}}
\left(\frac{\log{|f(r_m)|}}{r_m^{d}}
+\frac{\log{|f(r_{\epsilon})|}-\log{|f(r_{u_m})|}-\log{|f(r_{v_m})|}}{r_{h}^{d}}
+\frac{\log\frac{\alpha\beta r_{h}^2}{L^2}}{r_{h}^{d}}
-\frac{\log\frac{\alpha\beta r_m^2}{L^2|}}{r_m^{d}}\right).\nonumber
\eea
Here we have used the fact that $\{r_{u_m},r_{v_m},r_{\epsilon}\}\approx r_h$. On the 
other hand by making use of the fact that \cite{Agon:2018zso}
\be\label{Hlim}
\log |f(r_{u,v})|=\log|uv|+ c_0+{\cal O}(uv)\;\;\;\;\;\;\;\;\;\;\;{\rm for}\;\;uv\rightarrow 0,
\ee
 one gets
\bea
\log |f(r_{u_m})|&=&\log|u_m\epsilon_v|+ c_0+{\cal O}(\epsilon_v),\cr &&\cr
\log |f(r_{v_m})|&=&\log|\epsilon_u v_m|+ c_0+{\cal O}(\epsilon_u),\cr &&\cr
\log |f(r_{\epsilon})|&=&\log|\epsilon_u\epsilon_v|+ c_0+{\cal O}(\epsilon_u\epsilon_v),
\eea
which can be used to simplify \eqref{joint-2}  as  follows
\be
I^{\rm joint}_{\rm PI}=
-\frac{V_d L^d}{8\pi G_{N}}
\left(\frac{\log{|f(r_m)|}}{r_m^{d}}
-\frac{\log (u_m v_m)+c_0}{r_{h}^{d}}
+\frac{\log\frac{\alpha\beta r_{h}^2}{L^2}}{r_{h}^{d}}
-\frac{\log\frac{\alpha\beta r_m^2}{L^2}}{r_m^{d}}\right).
\ee
Here  $c_0=\psi^{(0)}(1)-\psi^{(0)}(\frac{1}{d+1})$ is a 
positive number and $\psi^{(0)}(x)=\frac{\Gamma'(x)}{\Gamma(x)}$ is the 
digamma function.

Finally one has to remove the ambiguity due to the normalization of the null vectors 
by adding the extra term \eqref{AMB}
to the action. The resulting expression is then
\be
I^{\rm amb}_{\rm PI}=
-\frac{V_dL^d}{8\pi G_N}\left(\frac{\log\frac{\alpha\beta \tilde{L}^2 r_m^2}{{L}^4}}
{r_m^{d}}+
\frac{2}
{d\,r_m^{d}}\right)+\frac{V_dL^d}{8\pi G_N}\left(\frac{\log\frac{\alpha\beta \tilde{L}^2r_h^2}{L^4}}
{r_h^{d}}+
\frac{2}
{d\,r_h^{d}}\right).
\ee
Therefore altogether for the subregion given by the intersection of the  WDW patch with the 
past interior  shown in the left panel of 
figure \ref{fig:E} one gets 
\be
I_{\rm PI}=\frac{V_d L^d}{8\pi G_{N}}
\left(\frac{1}{r_h^{d}}\log\frac{\tilde{L}^2}{L^2}+\frac{c_0}{r_{h}^{d}}
-\frac{(d+1)\tau}{2r_{h}^{d+1}}-\frac{\log\frac{\tilde{L}^2|f(r_m)|}{L^2}}{r_m^{d}}
\right),
\ee
which depends on time trough its $r_m$ dependence, as expected. Here we have used 
the fact  that $\log (u_m v_m)=-f'(r_h)r^*(r_m)=-\frac{(d+1)\tau}{2r_h}$.  
Note that for $r_m\rightarrow r_{\rm Max}$ 
the time dependence of the on-shell action drops out resulting to
 \be
 I_{\rm PI}=
\left(c_0-\frac{(d+1)\tau_c}{2r_h}+\log\frac{\tilde{L}^2}{L^2}\right)\frac{S_{th}}{2\pi}.
\ee
Note also that at late times where $r_m\rightarrow r_h$, using equation \eqref{Hlim},
the total on-shell action in the past interior vanishes.

\subsubsection{Future Interior}

Let us now compute on-shell action for the intersection of the WDW patch with the  
future interior shown in the right panel of figure \ref{fig:E}. In this case, using our previous 
results, the bulk term of the action is
\bea
I^{\rm bulk}_{\rm FI}=-\frac{V_{d}L^d}{4\pi G_N}(d+1)\int_{r_h}^{r_{Max}} \frac{dr}{r^{d+2}}
\left(\frac{\tau}{2}+r^*(\epsilon)-r^*(r)\right)
=-\frac{V_{d}L^d}{4\pi G_N}\left(\frac{1}{d\,r_h^d}
+\frac{\tau+\tau_c}{2r_h^{d+1}}\right).
\eea
There are five joint points two of which  have zero contributions for large $r_{Max}$, while 
the contributions of other three  points are given by
\bea\label{joint-3} 
I^{\rm joint}_{\rm FI}&=&\frac{V_d L^d}{8\pi G_{N}}
\left(\frac{\log\frac{\alpha\beta r_\epsilon^2}{L^2|f(r_\epsilon)|}}{r_\epsilon^{d}}
-\frac{\log\frac{\alpha\beta r_{u_{m'}}^2}{L^2|f(r_{u_{m'}})|}}{r_{u_{m'}}^{d}}
-\frac{\log\frac{\alpha\beta r_{v_{m'}}^2}{L^2|f(r_{v_{m'}})|}}{r_{v_{m'}}^{d}}
\right)\\&&\cr
&=&-\frac{V_d L^d}{8\pi G_{N}}
\left(\frac{\log{|f(r_{\epsilon})|}-\log{|f(r_{u_{m'}})|}-\log{|f(r_{v_{m'}})|}}{r_{h}^{d}}
+\frac{\log\frac{\alpha\beta r_{h}^2}{L^2}}{r_{h}^{d}}
\right)\cr &&
\cr
&=&\frac{V_d L^d}{8\pi G_{N}}
\left(\frac{\log |u_{m'} v_{m'}|+c_0}{r_{h}^{d}}
-\frac{\log\frac{\alpha\beta r_{h}^2}{L^2}}{r_{h}^{d}}
\right).\nonumber
\eea
The boundary terms associated with the null boundaries  vanish using affine parametrization 
for the null directions and the only term we need to compute is the surface term at
 future singularity. This is indeed the term we have already computed in \eqref{SURF}
 \be
I^{\rm surf}_{\rm FI}=\frac{V_dL^d}{8\pi G_N }(d+1)\,\frac{\tau+\tau_c}{2r_h^{d+1}}\,.
\ee
The only remaining contribution to be computed is the term needed to remove the ambiguity 
\be
I^{\rm amb}_{\rm FI}=\frac{V_dL^d}{8\pi G_N}\left(\frac{\log\frac{\alpha\beta \tilde{L}^2r_h^2}
{L^4}}
{r_h^{d}}+
\frac{2}
{d\,r_h^{d}}\right).
\ee
Taking all terms into account we have
\be
I_{\rm FI}=\frac{V_d L^d}{8\pi G_{N}}
\left(\frac{d\,\tau}{r_h^{d+1}}+\frac{(d-1)\tau_c}{2r_h^{d+1}}+\frac{c_0}{r_{h}^{d}}
+\frac{1}{r_{h}^{d}}\log\frac{\tilde{L}^2}{L^2}\right).
\ee
Here to get the final result we have used the fact that  $\log |u_{m'}v_{m'}|=\frac{(d+1)\tau}{2r_h}$.

It is also interesting to sum the contributions of both regions shown in figure \ref{fig:E} and
compare the resultant expression with the on-shell action evaluated on the whole WDW patch 
\be
I_{\rm Ext}=I_{\rm WDW}-(I_{\rm PI}+I_{\rm FI})=2\times
\frac{V_dL^d}{8\pi G_N}\bigg[
-\frac{c_0}{r_{h}^{d}}+
\left(\frac{1}
{\epsilon^{d}}-\frac{1}
{r_h^{d}}\right)\log\frac{\tilde{L}^2}{L^2}\bigg].
\ee
that is time  independent, as expected.
In fact this is the contribution of the part of the WDW patch that is outside of the black
hole horizon. The factor of two is a symmetric factor between left and right sides 
of the corresponding WDW patch.  It is also interesting to note that 
the finite term is negative! We will consider the above result in the next section where 
we will study subregion complexity.


\subsection{Late Time Behavior}

In this section we will study the time derivative of the on-shell actions we have found in 
the previous subsections. To proceed we note that from definitions of $r^*$ and $r_m$ 
one has 
\be
\frac{dr^*(r_m)}{d\tau}=-\frac{1}{2},\;\;\;\;\;\;\;\;\;\;\frac{dr_m}{d\tau}=\frac{1}{2}
f(r_m),
\ee 
which can be used to show
\be\label{TA}
\frac{dI_{\rm WDW}}{d\tau}=\frac{dI_{\rm past}}{d\tau}
=2M\left(1+\frac{1}{2} \tilde{f}(r_m)\log\frac{\tilde{L}^2|f(r_m)|}{L^2}\right),\;\;\;\;\;\;\;\;
\tilde{f}=\frac{r_h^{d+1}}{r_m^{d+1}}-1.
\ee
It is also interesting to compute the time derivative of the on-shell action for the individual 
subregions we have considered before. Actually it is straightforward to see
\be
\frac{dI_{\rm  PI}}{d\tau}=M\tilde{f}(r_m)\log\frac{\tilde{L}^2|f(r_m)|}{L^2},
\;\;\;\;\;\;\;\frac{dI_{\rm  FI}}{d\tau}=2M,\;\;\;\;\;\;\;
\frac{dI_{\rm  Ext}}{d\tau}=0\,.
\ee
It is evident that summing up these contributions one gets \eqref{TA}, as expected. Note that 
at late times where $r_m\rightarrow r_h$, the past interior has no contribution to the rate of complexity growth.


 Of course it is known that the complexity obtained from WDW patch violates the Lloyd's bound,
though at late time it approaches $2M$. From the above results it is evident that the contribution 
to the late time behavior comes from the  future interior of the black brane. It is also worth 
noting  that the violation of the Lloyd's bound is due to the contribution of the joint point 
located at the past interior. This, in turns, suggests that if one defines the complexity as 
on-shell action on the intersection of WDW patch and future interior the resultant 
complexity fulfills the bound and has linear growth all the time!  Of course if one wants
 appropriate UV divergences before regularizing the complexity we should also add 
 the contributions of the exterior region too.  More
explicitly one has
\bea
\tilde{I}_{\rm WDW}=I_{FI}+I_{\rm Ext}=2M
\left(\tau+\frac{(d-1)\tau_c-2c_0 r_h}{d}
+\left(\frac{2r_h^{d+1}}{d\,\epsilon^{d}}-\frac{r_{h}}{d}\right)\log\frac{\tilde{L}^2}{L^2}\right).
\eea
 
It is also instructive to note that at late time where  $r_m\rightarrow r_h$, setting 
$r_m-r_h=\xi$, from equation \eqref{rm} one finds 
\be
\tau=-\frac{2r_h}{d+1}\left(\log\frac{(d+1)\xi}{r_h}-c_0\right)\sim \frac{\beta}{2\pi} \log
\frac{r_h}{(d+1)\xi}\,,\;\;\;\;\;\;{\rm with}\;\beta=\frac{1}{T}\,.
\ee
In particular when one is away from the horizon about
a power of Planck scale $\xi\sim \frac{\ell_p^d}{r_h^{d-1}}$ the above late time behavior reads
\be
\tau\sim \frac{\beta}{2\pi}\log S_{th},
\ee
in which the on-shell action reads  $I \sim S_{th} \log S_{th}$, that is the scrambling complexity. 
Following \cite{Brown:2017jil} one may also consider the case where the time is about of
$\tau \sim\frac{\beta}{2\pi}\, e^{S_{th}}$ that could be the 
time where one gets maximum complexity. At that time the on-shell action is 
\be
I\sim S_{th}\, e^{S_{th}},
\ee
which could be thought of maximum complexity of the system\cite{Brown:2017jil}.


\subsection{Holographic Uncomplexity}  

\begin{figure}
\begin{center}
\includegraphics[scale=0.8]{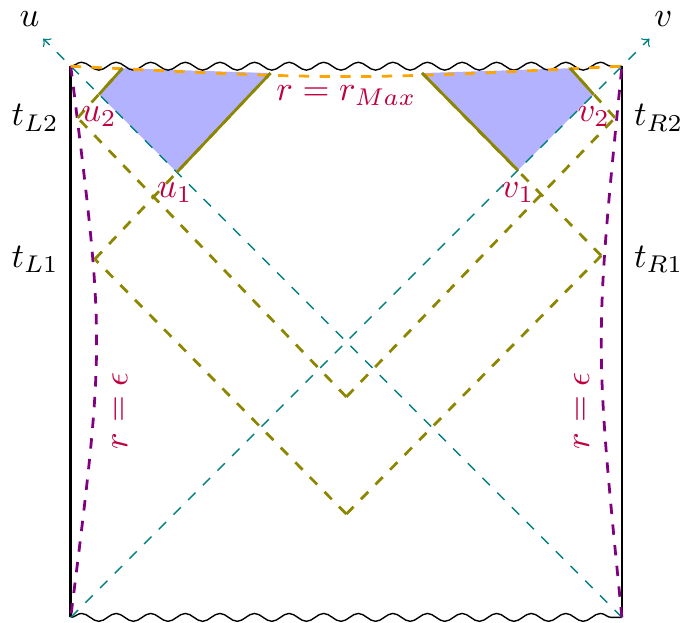}
\end{center}
\caption{Spacetime region corresponding to evaluation of holographic uncomplexity. The 
regions shown by blue color compute  uncomplexity given by the equation \eqref{UC}. This 
is not equal to difference between maximum complexity and the complexity of the 
state at a given time (see equation \eqref{UU}).}
\label{fig:Aun}
\end{figure}

Given a time slice and the associated WDW patch one may want to compute on-shell action 
on a region that should be included in the WDW patch as  time goes. 
The corresponding region is shown in figure \ref{fig:Aun}. Actually 
following \cite{Brown:2017jil} one may identify the on-shell action on this region with 
 ``holographic uncomplexity'' that is the gap between the complexity and the maximum possible 
 complexity (see also \cite{{Zhao:2017isy},{Stoltenberg:2018ink}}). In other words the 
 uncomplexity is a room for complexity to increase.  Alternatively one could
 thought of the  holographic uncomplexity as 
 the  spacetime resource available to an observer who intends to enter 
 the horizon\cite{Brown:2017jil}. 

Clearly the on-shell action on the region depicted in figure \ref{fig:Aun} 
is given by a difference of on-shell action evaluated on the future interior 
 \be\label{UC}
I_{\rm UC}=I_{\rm FI\,2}-I_{\rm FI\,1}=2M ({\tau}_2-\tau_1)\,.
\ee
where $\tau$ is the actual boundary time. 
It is also important to note that $\tau_2$ 
should be thought of a time cut off  and eventually we are interested in 
$\tau_2\rightarrow \infty$ limit for some fixed $\tau_1$. Indeed the time 
cut off could be set to $\tau_2\sim \frac{\beta}{2\pi} e^{S_{th}}$.

As we mentioned the holographic uncomplexity is defined as a difference 
between maximum complexity and the complexity of the state at a given time, it is then 
evident from \eqref{UC} that this equation  can not capture this difference. The crucial point 
is that the complexity, as we already mentioned, has two components: one from the boundary 
term and one from the joint point. The resultant uncomplexity given in equation \eqref{UC} does 
not fully contain the contribution of joint point. To be precise using equation \eqref{com} one has
\bea
\Delta I_{\rm WDW}&=&I_2^{\rm WDW}-I_1^{\rm WDW}\cr &&\cr &=&
\frac{V_dL^d}{8\pi G_N}\left[\frac{d-1}
{2r_h^{d+1}}(\tau_2-\tau_1)-
\frac{\log |f(r_{m\,2})|}{r_{m\, 2}^d}+\frac{\log |f(r_{m\,1})|}{r_{m\, 1}^d}\right].
\eea
 One observes that there is a joint contribution that the subregion shown 
in figure \ref{fig:Aun} can not see it and thus it is not equal to $I_{\rm UC}$.  
Of course it approaches $I_{\rm UC}$ when both $r_{m\,1}$ and $r_{m\,2}$
approach the horizon. Actually using the fact that $\tau_2$ should be thought of 
a cut off and therefore it is large ({\it i.e.} $r_{m\,2}\rightarrow r_h$)  the above expression reads
\bea\label{UU}
\Delta I_{\rm WDW}\approx 2M(\tau_2-\tau_1)-
\frac{V_dL^d}{8\pi G_N}\left[\frac{c_0}{r_h^d}-\frac{(d+1)\tau_1}
{2r_h^{d+1}}-\frac{\log |f(r_{m\,1})|}{r_{m\, 1}^d}\right].
\eea
Note that the second part is just the on-shell action evaluated on the past interior that vanishes 
as $r_{m\, 1}$ approaches the horizon. It is worth mentioning  that, although it is not clear from this expression,  setting $\tau_1=\tau_2$ the above equation vanishes. To see this we note that 
in this expression the time  $\tau_2$ is associated with the limit 
where $r_{m\, 2}\approx r_h$. Therefore  setting $\tau_1 =\tau_2$ the joint 
point $r_{m\,1}$ approaches the horizon too and thus  the expression in the bracket 
vanishes in this limit as well.


\section{Subregion Complexity and Outside  the Horizon }\label{sec:2sided}

In the previous section we have computed on-shell action on different regions 
containing a part that is located behind the horizon. These regions could be 
found by the intersection of a WDW patch with the interior of a two sided  black brane.
We have seen that the resultant on shell action is time dependent whenever it  receives 
a contribution from a region inside the black brane. 

On the other hand one may consider  cases where the subregions of interest are entirely 
outside  the horizon. This is, indeed, what we would like to consider in this section. 
 In this case, unlike the previous cases, one usually gets time independent 
on shell action. In some case the region of interest could be thought of as intersection of 
the WDW patch with entanglement wedge.

\subsection{Complexity of Layered Stretched Horizon }

\begin{figure}
\begin{center}
\includegraphics[scale=0.9]{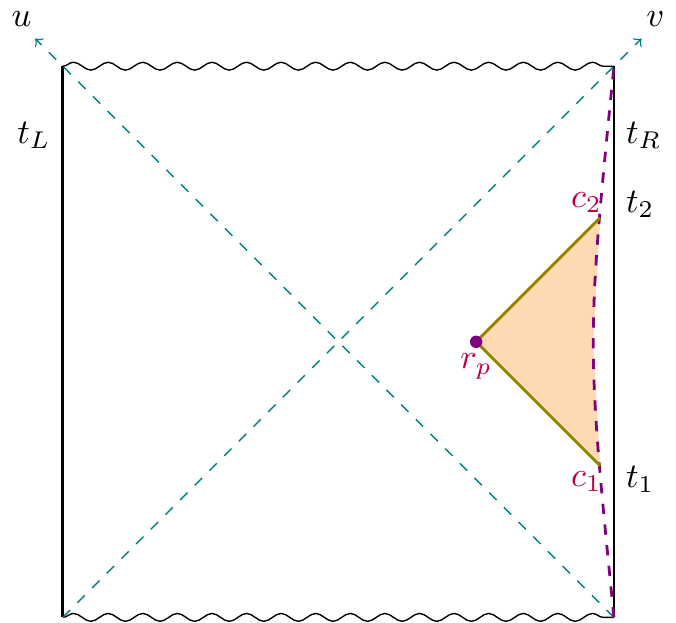}
\includegraphics[scale=0.9]{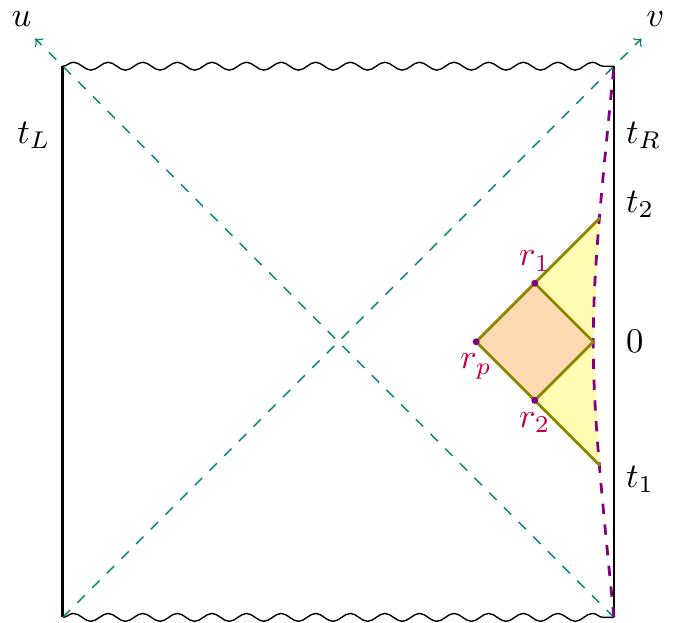}
\end{center}
\caption{{\it Left:} A localized operator at $P$. The colored region is the part that is involved in the 
construction of the operator localized  at $r=r_p$. {\it Right:} The orange region is the 
intersection of WDW patch and entanglement wedge at time slice $t_R=0$ for half of 
an eternal black hole.}
\label{fig:A}
\end{figure}

Let us consider a subregion in the black hole exterior in the shape of a triangle  shown 
in the left panel of  figure \ref{fig:A}. The three faces of the corresponding  triangle are given by 
two null and a  timelike boundaries 
 \bea\label{NBR}
 t=t_1+r^*(\epsilon)-r^*(r),\;\;\;\;\;\;\;\;\;\;\;t={t_2}-r^*(\epsilon)+r^*(r),\;\;\;\;\;\;\;\;\;\;r=\epsilon\,.
 \eea
The null boundaries intersect at the point $r=r_p$ that is given by 
 \be\label{rp}
 \tilde{\tau}\equiv t_{R\,2}-t_{R\, 1}=2\left(r^*(\epsilon)-r^*(r_p)\right).
 \ee
 where  $\tilde{\tau}$ is the time interval. This  should not be confused with  the actual field theory 
 time coordinate $\tau$ we have used in the previous section.  
 
Actually  following \cite{Susskind:2014rva} where the author has considered layered 
stretched horizon, this might be thought of as  a bulk operator $P$ localized at 
point $r_p$. Indeed the corresponding  triangle  shows a region of the boundary involved 
in the construction of the operator $P$.  Now the aim is to compute the on-shell action 
in this subregion. Following \cite{Susskind:2014rva} the result might be thought of as 
complexity of the operator localized on the corresponding layer.

To proceed  let us start with the bulk contribution. From the notation depicted in 
figure \ref{fig:A} it is straightforward to see
\bea
 I^{\rm bulk}_{\rm Tri}&=&-\frac{V_dL^d}{8\pi G_N}(d+1)\int^{r_p}_{\epsilon} 
\frac{dr}{r^{d+2}} \left(\tilde{\tau}-2({r^*(\epsilon)-r^*(r)})\right)\cr &&\cr
&=&-\frac{V_d L^d}{8\pi G_N}\left(\frac{1}{\epsilon^{d+1}}-\frac{1}{r_h^{d+1}}\right)\tilde{\tau}+\frac{V_dL^d}{4\pi G_Nd}\left(\frac{1}{\epsilon^d}-\frac{1}{ r_p^d}\right).
\eea
As for the boundary terms  we only need  to consider  the Gibbons-Hawking-York term at  
the timelike boundary  $r=\epsilon$
 \be
I^{\rm surface}_{\rm Tri}=\frac{1}{8\pi G_N}\int d^dx\,
 \int_{t_1+r^*(\epsilon)-r^*(r)}^{t_2-r^*(\epsilon)+r^*(r)} dt\, \sqrt{h}K_t 
\Big|_{r=\epsilon} = \frac{(d+1)V_dL^d}{8\pi G_N }\,
\left(\frac{1}{\epsilon^{d+1}}-\frac{1}{2r_h^{d+1}}\right)
\tilde{\tau}.
\ee
The normal vectors associated with the boundaries of the triangle given by \eqref{NBR} are
 \be
n_t=\frac{Ldr}{\epsilon\sqrt{f(\epsilon)}},\;\;\;\;\;\;
k_1={\alpha}\left(-dt+
\frac{dr}{f(r)}\right),\;\;
\;\;\;\;k_2={\beta}\left(dt+
\frac{dr}{f(r)}
\right)\,,
\ee
which can be used to compute the contribution of the joint points as follows  
\bea
I^{\rm joint}_{\rm Tri}&=&
\frac{1}{8\pi G_N}\int_{c_1} d^dx \sqrt{\gamma} 
\;\log \left|{k_2\cdot n_t}\right|
+\frac{1}{8\pi G_N}\int_{c_2} d^dx \sqrt{\gamma} \;\log \left|{k_1\cdot n_t}\right|\cr
 &&\cr
&&-\frac{1}{8\pi G_N}\int_{r_p} d^dx \sqrt{\gamma} \;\log \left|\frac{k_1\cdot k_2}{2}\right|\cr 
&&\cr
&=&\frac{V_dL^d}{8\pi G_N}\,\frac{\log{\frac{\alpha\beta\epsilon^2}{L^2}}}{\epsilon^d}
-\frac{V_dL^d}{8\pi G_N}\left(\frac{
\log {\frac{\alpha\beta r_p^2}{L^2}}}{r_p^d}-\frac{\log f(r_p)}{r_p^d}\right).
\eea
The contribution of the term needed to  remove the ambiguity is
\bea
I^{\rm amb}_{\rm Tri}&=&\frac{1}{8\pi G_N}\int_{\rm null} d\lambda d^dx \sqrt{\gamma}
\Theta\log\frac{|\tilde{L} \Theta|}{d}\cr &&\cr
&=&
\frac{V_dL^d}{8\pi G_N}\left(\frac{\log\frac{\alpha\beta\tilde{L}^2 r_p^2}{L^4}}
{r_p^{d}}+
\frac{2}
{d\,r_p^{d}}\right)-\frac{V_dL^d}{8\pi G_N}
\left(\frac{\log\frac{\alpha\beta\tilde{L}^2 \epsilon^2}{L^4}}
{\epsilon^{d}}+
\frac{2}
{d\,\epsilon^{d}}\right).
\eea
Therefore taking all contributions into account one arrives at
\be\label{TT}
I_{\rm Tri}=\frac{V_d L^d}{8\pi G_N}\left[\left(\frac{d}{\epsilon^{d+1}}
-\frac{d-1}{2r_h^{d+1}}\right)\tilde{\tau}+\left(\frac{1}
{r_p^{d}}-\frac{1}
{\epsilon^{d}}\right)\log\frac{\tilde{L}^2 }{L^2}+\frac{\log f(r_p)}{r_p^d}\right].
\ee
At this stage we would like to recall that whenever one is dealing with the computation
of on shell action it is important to make it clear what one means by action. As we already
mentioned by an action we mean all terms needed to have a covariant action with a
well defined variational principle that result to a finite free energy. This, in particular,
requires to consider the  counter terms that obtained in the context of  holographic renormalization. In the present case where we have a  
timelike flat boundary at $r=\epsilon$, the corresponding   counterterm
is given by  
\be
I^{\rm ct}_{\rm Tri}=-\frac{1}{8\pi G_N}\int_{r=\epsilon} d^{d+1}x\,\sqrt{h}\, \frac{d}{L},
\ee
that gives the following contribution to the on shell action
\be
I^{\rm ct}_{\rm Tri}=
\frac{V_dL^d }{8\pi G_N}\left(-\frac{d}{\epsilon^{d+1}}+\frac{d}{2r_h^{d+1}}\right)\tilde{\tau}\,.
\ee
Combining the above result with equation \eqref{TT}, the total on-shell action reads\footnote{
We note that the resultant on shell action is still divergent due to the ambiguity of fixing the
length scale $\tilde{L}$. Of course there is a natural way to fix this length scale by assuming 
that the corresponding on shell action  for an  AdS geometry in the Poincare coordinates 
vanishes (as a reference state), leading to  ${\tilde L}=L$. Therefore one ends up with a
finite on shell action.}
\be\label{TT1}
I_{\rm Tri}=\frac{V_d L^d}{8\pi G_N}\left[-\frac{1}
{\epsilon^{d}}\log\frac{\tilde{L}^2 }{L^2}+\frac{\tilde{\tau}}{2r_h^{d+1}}
+\frac{\log \frac{\tilde{L}^2f(r_p)}{L^2}}{r_p^d}\right].
\ee 
Note that since we have already assumed $\tilde{L}\geq L$, from the above 
expression one finds that the most divergent term as well as the finite term are negative. 
This is of course in contrast with what one would expect from complexity.  Actually the 
result is reminiscent of free energy of the black hole. Indeed denoting  
the contribution of joint point by ${\cal J}$ one has (dropping the divergent term)
\be
I_{\rm Tri}=-{\cal F}\,\tilde{\tau}+{\cal J}_p\,\;\;\;\;\;\;\;{\rm with }\,\;\;\;{\cal J}_p=\frac{\log f(r_p)}
{r_p^d}\,,
\ee
where ${\cal F}=-\frac{V_d L^d}{16\pi G_N}\,\frac{1}{r_h^{d+1}}$ is the free energy of the 
corresponding black brane .  To summarize we note that the on-shell action in this case 
consists of two parts: the first part that might be thought of  as the classical contribution  is the 
contribution of the timelike boundary  that is the free energy of corresponding black brane, while
the second one that comes from joint point should be treated as the new contribution 
associated with the complexity of the operator.  Clearly when a given subregion does not 
contain a timelike boundary the free energy drops and the whole 
contributions come from joint points (see next subsection).

For the case where the point $r_p$ is in the 
vicinity of the horizon,  \textit{i.e.}, $r_p=r_h-\xi$ for 
 $\xi\ll r_h$, from equation \eqref{TT1} one finds
\be
I_{\rm Tri}\approx  \frac{V_d L^d}{8\pi G_N} \left(\frac{1}
{r_p^{d}}-\frac{1}
{\epsilon^{d}}\right)\log\frac{\tilde{L}^2 }{L^2}-\frac{V_d L^d}{16\pi G_N}\,\frac{1}{r_h^{d+1}}
(d\;\tilde{\tau}).
\ee
that shows the layer (operator) becomes more complex as one approach the horizon. In 
particular  when one is away from the horizon about the Planck length one gets 
\be
I_{\rm Tri}\approx  \frac{V_d L^d}{8\pi G_N} \left(\frac{1}
{r_p^{d}}-\frac{1}
{\epsilon^{d}}\right)\log\frac{\tilde{L}^2 }{L^2}-\frac{1}{2\pi (d+1)}\,S_{th}\log S_{th}\,
 \ee

\subsection{CA Proposal and Subregion Complexity}   
 
An immediate application of the result we have  obtained in the previous section is to 
find on-shell action for a square  subregion shown in orange in the right panel of figure 
\ref{fig:A}. The result may be used to compute on shell action on a region 
obtained by the intersection of entanglement wedge and WDW patch.
The desired result 
can be found by algebraic summation of three triangles identified by $r_1, r_2$ and 
$r_p$. Actually using equation \eqref{TT} one gets\footnote{ One could have 
directly computed the on-shell action for the square region taking into account 
all terms in the action. Of course the result is the same as what we have 
found by an algebraic summation of three triangles.}
\bea\label{ISq}
I_{\rm Sq}&=&I_{r_p}-I_{r_1}-I_{r_2}\\ &&\cr
&=&\frac{V_d L^d}{8\pi G_N}\bigg[\left(\frac{1}{\epsilon^d}+\frac{1}{r_p^d}-\frac{1}{r_1^d}-\frac{1}{r_2^d}\right)\log\frac{\tilde{L}^2 }{L^2}+\frac{\log f(r_p)}{r_p^d}-\frac{\log f(r_1)}{r_1^d}
-\frac{\log f(r_2)}{r_2^d}
\bigg].\nonumber
 \eea
Note that in this case the most divergent term is positive as expected 
for an expression representing complexity. Indeed since the corresponding 
subregion is the intersection of WDW patch and domain  of dependence of a subregion 
in the boundary  theory (which is the whole system in the present case at time $\tau=0$) 
we would like to identify this expression as the CA subregion complexity \cite{Carmi:2016wjl}.
Note that since there is  no timelike or spacelike
boundaries, all contributions come from the joint points.
 
\begin{figure}
\begin{center}
\includegraphics[scale=0.8]{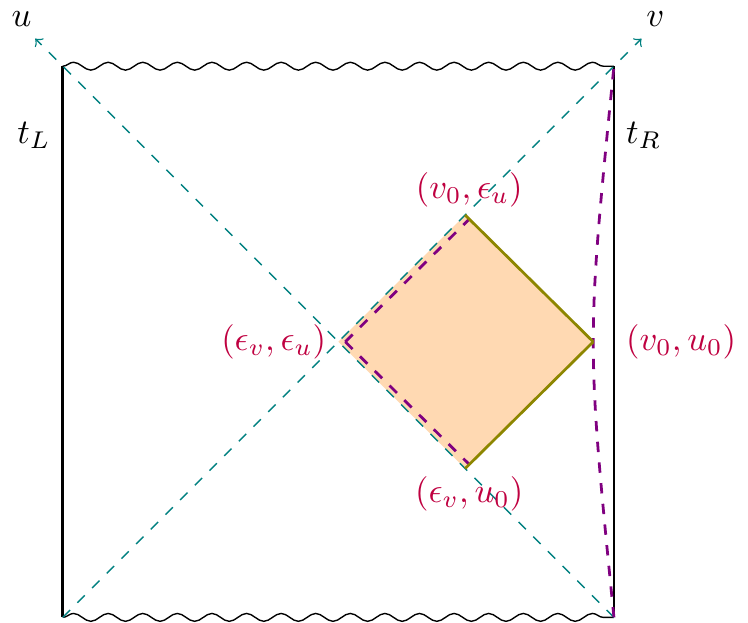}
\includegraphics[scale=0.8]{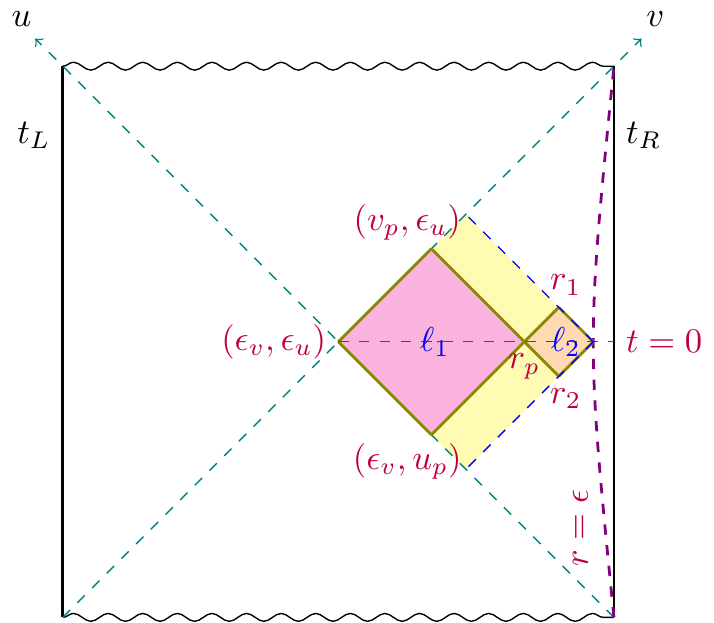}
\end{center}
\caption{{\it Left:} Intersection of WDW patch and entanglement wedge for 
large entangling region at time slice  $t_R=0$ for half of an eternal black hole.
{\it Right:} Two subregions denoted by $\ell_1$ and $\ell_2$. }
\label{fig:D}
\end{figure}

It is also interesting to consider the limit of  $\{r_p, r_1, r_2\}\rightarrow r_h$, where 
we get a subregion shown in the left panel of figure \ref{fig:D}. This is the intersection of WDW patch
at time slice $\tau=0$  with the right exterior of the black hole.
Actually by making use of equation \eqref{Hlim} and with the notation shown in figure \ref{fig:D} in
the limit of  $\{r_p, r_1, r_2\}\rightarrow r_h$, equation \eqref{ISq} reads
\be\label{SUB}
I_{\rm Sq}=\frac{V_d L^d}{8\pi G_N}\left(\frac{1}{\epsilon^d}-\frac{1}{r_h^d}\right)\log
\frac{\tilde{L}^2 }{L^2}-\frac{c_0}{2\pi}\,S_{th}\,.
\ee
 It is important to note that although the 
most divergent term is positive for $\tilde{L}> L$, the finite term is negative.  We note
that on-shell action for the subregion shown in the left panel of figure \ref{fig:D} has 
been recently studied 
 \cite{Agon:2018zso} where the authors have not considered the term needed to remove the 
 ambiguity and have fixed the ambiguity by hand\footnote{We note that the contribution of the
 term removing the ambiguity has been added to the published version of \cite{Agon:2018zso}
 where a proper citation to the present paper is also given.}. As a result the finite term they have found was 
 positive.  We note, however, that it is crucial to take into account 
 the corresponding term to maintain the reparamerization  invariance of the action. 
Note that for values of $r_p$ it can be seen that the finite part of equation
\eqref{ISq} is always negative\footnote{We would like to thank B. Swingle for 
discussions on this point.}. It is also interesting to compare on-shell action evaluated on 
different subregions and union 
of the subregions. To proceed we will consider two subregions denoted by $\ell_1$ and $\ell_2$
in the right panel of figure \ref{fig:D}. Using the notation shown in the figure and 
setting  $\tilde{L}=L$, one has
\bea
I_{\ell_1}&=&\frac{V_dL^d}{8\pi G_N}\left(\frac{\log{|f(r_p)|}}{r_p^{d}}
+\frac{\log{|f(r_{\epsilon})|}}{r_{h}^{d}}
-\frac{\log{|f(r_{u_p})|}}{r_{h}^{d}}
-\frac{\log{|f(r_{u_p})|}}{r_{h}^{d}}
\right)\cr &&\cr
&=&\frac{V_dL^d}{8\pi G_N}\left(\frac{\log{|f(r_p)|}}{r_p^{d}}
-\frac{c_0+\log|u_pv_p|}{r_{h}^{d}}
\right)=\frac{V_dL^d}{8\pi G_N}\left(\frac{\log{|f(r_p)|}}{r_p^{d}}
-\frac{c_0-f'(r_h)r^*(r_p)}{r_{h}^{d}}
\right),
\cr &&\cr
I_{\ell_2}&=&\frac{V_dL^d}{8\pi G_N}\left(\frac{\log |f(r_p)|}{r_p^d}-\frac{\log |f(r_1)|}{r_1^d}
-\frac{\log |f(r_2)|}{r_2^d}\right)\,.
\eea
Here in order to simplify $I_{\ell_1}$ we have used \eqref{Hlim}. On the other 
hand the on-shell action evaluated on $\ell_1\cup \ell_2$ is \eqref{SUB}
\be
I_{\ell_1\cup\ell_2}=-\frac{V_dL^d}{8\pi G_N}
\frac{c_0}{r_{h}^{d}}\,.
\ee
Therefore one gets
\bea\label{A}
A\equiv I_{\ell_1}+I_{\ell_2}-I_{\ell_1\cup\ell_2}=
\frac{V_dL^d}{8\pi G_N}\left(2\frac{\log{|f(r_p)|}}{r_p^{d}}
-\frac{(d+1) r^*(r_p)}{r_{h}^{d+1}}
-\frac{\log |f(r_1)|}{r_1^d}
-\frac{\log |f(r_2)|}{r_2^d}\right)\,.
\eea
It is then important to determine  the sign of $A$. To do so, one first observes that $A$ vanishes
at both $\{r_p,r_1,r_2\}\rightarrow r_h$ and $\{r_p,r_1,r_2\}\rightarrow 0$ limits. On the 
other hand one can show that at $\{r_p,r_1,r_2\}\approx 0$ the function $A$ approaches zero
from above leading to the fact that $A\geq 0$. This behavior may also be shown numerically. As a result we conclude that the on-shell action we have evaluated  
for subregions in the exterior of the black brane obeys  subadditivity condition
\be
I_{\ell_1}+I_{\ell_2}\geq I_{\ell_1\cup\ell_2}\,,
\ee
that
is indeed in agreement with results of \cite{Agon:2018zso} (see also \cite{Camargo:2018eof}). 
It is worth noting that in order 
to reach the above result,  the contribution of the corner term,  $\log u_pv_p$, plays a crucial
role.

 
\section{Discussions and Conclusions}

In this paper motivated by ``complexity equals action'' proposal we have 
evaluated  on-shell action on certain spacetime subregions enclosed by 
null boundaries that of course includes  the WDW patch itself too. Our main 
concern was to compute finite term of the on-shell action. It is contrary to the most 
studies in the literature where the main concern was to compute the growth rate of 
the complexity (see for example \cite{{Momeni:2016ira},{Alishahiha:2017hwg},
{Nagasaki:2017kqe},{Miao:2017quj},{Ghodrati:2017roz},{Qaemmaqami:2017lzs},
{Sebastiani:2017rxr},{Cano:2018aqi},{An:2018dbz},{Fareghbal:2018ngr},{Ghodrati:2018hss},
{Zhang:2018qnt},{Mahapatra:2018gig},{Tanhayi:2018gcj}}).

Although we have computed on-shell action on a given subregion, taking into 
account all terms needed to have reparametrization invariance and well-defined
variational principle, we have observed that the final result is  given by contributions of
joint points and  timelike or spacelike boundaries. 
Removing the most divergent term by setting $\tilde{L}=L$, the corresponding 
joint contribution, ${\cal J}$, and timelike and spacelike surface contributions, 
 ${\cal S}_t$, and ${\cal S}_s$ are given by
\be
{\cal J}=\frac{V_dL^d}{8\pi G_N}\frac{\log |f(r)|}{r^d},\;\;\;\;\;\;\;\;\;\;\;\;
{\cal S}_t=\frac{V_dL^d}{8\pi G_N}\frac{\tilde{\tau}}{2 r_h^{d+1}},\;\;\;\;\;\;\;\;\;\;
{\cal S}_s=\frac{V_dL^d}{8\pi G_N}\frac{(d-1)\tau}{2 r_h^{d+1}}\,.
\ee
Note that when the joint point occurs at the horizon one needs to take $r\rightarrow r_h$ 
limit from the above joint contribution ${\cal J}$ that typically results to an expression 
proportional to $\log|u v|+c_0$.

 Clearly when the joint point is located at the horizon one needs to regularize the 
joint contribution using equation \eqref{Hlim}. The sign of the joint contribution
depends on the position of the corresponding joint point. 
If the joint point locates on the left or right of the
given subregion, the sign is positive and for those that located up or down of the subregion, it is 
negative. It is also interesting to compute time derivative of the above expressions 
\be
\dot{{\cal J}}=\frac{V_dL^d}{8\pi G_N}\left(\frac{d+1}{2r_h^{d+1}}+\frac{d}{2r^{d}}f(r)
\log |f(r)|\right),\;\;\;\;\;\;\;\;\;\;\;\;
\dot{{\cal S}}_t=0,\;\;\;\;\;\;\;\;\;\;
\dot{{\cal S}}_s=\frac{V_dL^d}{8\pi G_N}\frac{(d-1)}{2 r_h^{d+1}}\,,
\ee
showing that in the late time the joint point has nontrivial contribution. 

Another observation we have made is that whenever the subregion contains a part of
black hole interior the finite part of the action is positive and time dependent, while for 
the cases that the desired subregion is entirely  in the exterior part of the black hole 
the corresponding  finite term is time independent and negative. It is also important to 
note that for all cases, except one, the most divergent term exhibits  volume law scaling with 
positive sign. These points should be taken into account when  it comes to interpret the 
results from field theory point of view.

Throughout this paper we have been careful enough to clarify what we mean by 
on-shell action.  Indeed there are several terms one may add to action that could 
alter the results once we compute the on-shell action. It is then important to fix them.
Our physical constraints were to have reparametrization invariance and a well-defined 
variational principle. These assumptions enforce us to have certain boundary and
joint actions. In particular it was crucial to consider the log term given by equation \eqref{AMB} that is needed  
to remove the ambiguity associated with the null vectors. Actually in our computations this 
term has played an essential role.

We note, however, that even with this term the resultant on-shell action still contains
an arbitrary length scale. We have chosen the length scale so that the most divergent 
term of the on-shell action is positive. This is indeed required if one wants to identify 
the on-shell action with complexity, at least when evaluated inside the WDW patch.
Of course following the general idea of the holographic renormalization one may add certain 
counterterms to remove all divergent terms including that associated with the undetermined 
length scale\cite{TBA}. This is, actually,  what we have done in this paper when we were
 only interested in the finite part of the on-shell action.  

In fact if one wants to identify the on-shell action with the complexity we may not be 
surprised to have an arbitrary length scale. This might be related to choosing an arbitrary length 
scale in the definition of complexity in quantum field theory (see {\it e.g.} 
 \cite{Jefferson:2017sdb,Chapman:2017rqy}). Of course eventually  we would like to 
 find a way to fix the length scale or at least to make a constraint on it  so that it could
 naturally lead to a clear interpretation in terms of complexity.   
 
The main question remains to be addressed is how to interpret our results from field theory 
point of view. It is well accepted that the on-shell action evaluated in the WDW patch is
associated with the computational complexity that is the minimum number of gates one needs
to reach the desired  target state from reference state (usually the vacuum state). Of course this 
is the complexity defined for a  pure state. It is then natural to look for a definition of complexity 
for mixed state. This has been partially addressed in   \cite{{Agon:2018zso},
{Camargo:2018eof}}.

When we are restricted to a subregion,  even the whole system is in a pure state,   
we would have a mixed state density matrix  and therefore a definition of mixed state 
complexity is required. Of course the resultant  subregion complexity could as well 
depend on the state of whole system whether or not it is pure.

Actually different possible definitions of subregion complexity have been explored in 
\cite{Agon:2018zso}. Based on  results we have found it seems that the on-shell action 
evaluated on a subregion in the exterior is better match with purification complexity
that is the pure state complexity of a purified state minimizing over all possible purifications.
The main observation supporting this proposal is the subadditivity condition satisfied by the 
corresponding on-shell actions\footnote{We would like to thank B. Swingle for 
discussions on this point. }. Note that this is not the case for complexity obtained by 
CV proposal. 

Since we have been dealing with complexity for subregions, motivated by 
entanglement entropy and mutual information it might be useful to define 
a  new quantity associated with two subregions $A$ and $B$  as  follows
\be
{\cal C}(A|B)={\cal C}(A)+{\cal C}(B)-{\cal C}(A\cup B),
\ee
that could be thought of as mutual complexity. Here ${\cal C}$ 
stands for complexity evaluated using CA proposal. It also clear that the above
expression  that it is finite and  symmetric under the exchange of subregions A and B.

This quantity might be thought of as a quantum measure that measures  the correlation 
between  two subsystems. Of course we should admit that the precise
interpretation of this  quantity is not clear to us. Nonetheless it might be possible to explore 
certain properties of the quantity. In particular it is interesting to determine the sign of the mutual
complexity which in turns could tell us whether the complexity is superadditive  or subadditive.

Actually with the  context we have been studying the complexity in this paper,  it is clear that 
if one considers two subregions from two different boundaries  of the eternal black brane 
(one from left and the other from right boundaries) the above quantity is negative.
More precisely in this case the  mutual complexity is given in terms of the on shell action 
evaluated in the past and future interiors multiplied by a minus sign (see, for example,
subsection 2.3.2). 
Of course, note that in this case since we have to consider all regions inside the horizon, 
the final results are time dependent. 

On the other hand if we are interested in the time independent complexity, we 
will have to consider cases where both regions are located at the same boundary 
of the eternal black brane as shown in the Penrose diagram in figure \ref{fig:D}. 
In these cases, being time independent, we will consider a time 
slice at $\tau=0$. To explore the possible sign of the mutual complexity we first note that in 
general there is an arbitrary length scale, i.e. $\tilde{L}$, 
in the expression of the complexity  that could change the sign as one chooses different scales. This
scale appears due to the ambiguity of the normalization of the null  vectors (see equation 
\eqref{AMB}). 

In fact  in subsection 3.2 we have shown that  under the  assumption of $\tilde{L}=L$  
 the subregion complexity is subadditive that means the mutual complexity is positive. 
This is also consistent with the numerical result found in \cite{Camargo:2018eof}.
It is, however, important  to note that a priori there is no reason to fix the new scale 
$\tilde{L}$ equal to $L$ that is AdS radius. 

 Actually it is straightforward  to see that the sign of the mutual complexity defined above 
 also depends on the ratio $\frac{\tilde{L}}{L}$. More precisely from equation  
 \eqref{ISq} one observes  that there are terms proportional to $\log \frac{\tilde{L}^2}{L^2}$ 
 which  could contribute to the mutual complexity whose value is negative 
for $\frac{\tilde{L}}{L}>1$, Thus this makes the  mutual complexity negative. On the 
other hand for  $\frac{\tilde{L}}{L}\leq1$, their contributions are  positive that in turns make 
the mutual complexity positive too. 
 
 To conclude we note that the mutual complexity is positive for the cases where the subregion 
 complexity is time independent and  moreover  $\frac{\tilde{L}}{L}\leq1$. 
 It would be interesting to further explore properties of the  mutual complexity 
 which  could  be though of a quantity that diagnoses whether the complexity is sub or super 
 additive.

\subsection*{Acknowledgements}
The authors would like to kindly thank A. Akhavan, J. L. F. Barbon,  A. Naseh, F. Omidi, 
B. Swingle, M. R. Tanhayi and  M.H. Vahidinia for useful comments  and 
discussions  on related topics.  We would also like to thank the referee
for his/her useful comment.


\end{document}